\documentclass{article}
\usepackage{arxiv}
\usepackage{graphicx}
\usepackage{float}
\usepackage{amsmath,amssymb}
\usepackage{amsthm}
\usepackage{algorithm,algorithmic}
\usepackage[labelformat=simple]{subfig}

\usepackage{multirow}
\usepackage{latexsym,bm}
\usepackage{hyperref}
\usepackage{doi}     % for things like \Box and \Diamond

\graphicspath{{./fig/}}

\def\bx{\mathbf{x}}
\newtheorem{remark}{Remark}

\usepackage{hyperref}
\makeindex

\title{Structure preserving integration of 3D dissipative bi-Hamiltonian/Nambu systems}

\author{
  B\"ulent Karas\"ozen\\
     Department of Mathematics\\
     Middle East Technical University\\
     Ankara-Turkey\\
     \texttt{bulent@metu.edu.tr}\\
  \And
  Murat Uzunca\\
   Department of Mathematics\\ Sinop University\\
     Sinop-Turkey\\ \
     \texttt{muzunca@sinop.edu.tr}\\
}

\begin{document}

\maketitle

\begin{abstract}
A structure-preserving splitting integrator is developed for 3D dissipative bi-Hamiltonian/Nambu systems. The integrator uses Strang splitting for conservative and dissipative parts. For Nambu systems, the divergence-free, conservative part is integrated using the energy/volume-preserving Kahan's method, and the dissipative part is integrated by the forward and backward Euler methods. For dissipative bi-Hamiltonian systems, the conservative part is integrated with the energy-preserving average vector field (AVF) method. In both cases, the Hamiltonians of the conservative parts are preserved in the Lorenz, Chen, and Rabinovich systems. The periodic and chaotic solutions are computed accurately by the conservative-dissipative Strang splitting approach.
\end{abstract}

\keywords{Energy preservation, volume preservation, Poisson structure, splitting methods, dissipative systems}

{\bf \em MSC (2010)} 65P10, 037M15

%%%%%%%%%%%%%%%%%%%%%%%%%%%%%%%%%%%%%%%%%%%%%%%%%%%%%%%%%%%%%%%%%%%%%%%%%%%%%%%%%%%%%%%%%%
%%%%%%%%%%%%%%%%%%%%%%%%%%%%%%%%%%%%%%%%%%%%%%%%%%%%%%%%%%%%%%%%%%%%%%%%%%%%%%%%%%%%%%
\section{Introduction}

Three-dimensional (3D) systems such as the Lorenz, Chen, and Rabinovich systems can be recast as dissipative bi-Hamiltonian/Nambu systems \cite{Axenides10,Esen17,Katagiri26,Nevir94,Roupas12}. 
These equations consist of two parts: the non-dissipative conservative bi-Hamiltonian/Nambu part, and the dissipative gradient part. 
There exist two conserved integrals, the generalized or Nambu Hamiltonians $H_1$, $H_2$ in the non-dissipative part, which is referred to as the Nambu-Hamilton system \cite{Nambu73,Takhtajan94}. 
Nambu form represents a generalization of classical Hamiltonian mechanics, which is appropriate for the study of divergence-free, rotational phase-space volume preserving flows. 
They are integrable dynamical systems, with the trajectories in phase space given by the intersection of the surfaces defined by the conserved Nambu Hamiltonians $H_1$ and $H_2$.
The dissipative, irrotational part of the flow is the gradient of the dissipation function.
In this formulation, the velocity field separates into a reversible bi-Hamilton/Nambu component and an irreversible gradient component. 
The dynamics of the whole system are driven by the dissipative, irrotational part.
In three dimensions, these systems can also be considered as metriplectic systems, written in terms of the bi-Hamiltonian/Nambu bracket \cite{Esen17} with the two Hamiltonians.

The symplectic integrators have been developed by splitting the Hamiltonians and composing the time evolution \cite{Mclachlan02,McLachlan02s,Blanes24} to preserve modified, nearby Hamiltonians. 
Structure-preserving integrators are constructed by splitting the thermodynamic structure known as the general equation for the non-equilibrium reversible-irreversible coupling (generic) into reversible and irreversible dynamics \cite{Shang20}. 
However, these methods have been limited to preserving one particular conserved quantity, whereas Nambu systems have multiple conserved quantities, e.g., Hamiltonian or volume. 
Similarly, the bi-Hamiltonian systems have, besides the Hamiltonian, Casimirs as conserved quantities.
Splitting separable Nambu Hamiltonians, a structure-preserving composition method has been developed in \cite{Horikoshi24}, that preserves the shadow or modified Nambu Hamiltonians. Conservative-dissipative Strang splitting was applied to damped canonical Hamiltonian systems in \cite{Modin11,Viorel21}.
In this paper, we propose a structure-preserving integrator for dissipative bi-Hamiltonian/Nambu systems by splitting dissipative and non-dissipative parts. 
For Nambu systems, the divergence-free non-dissipative part is integrated by the volume-preserving linearly implicit Kahan's method \cite{Iserles15,Celledoni19,Celledoni24}, and the dissipative part by forward and backward Euler methods. 
The resulting conservative-dissipative Strang splitting is symmetric and second-order. 
Kahan's method is an integrable discretization for Nambu-Hamilton systems, i.e., the modified homogeneous and inhomogeneous quadratic Nambu Hamiltonians are preserved.\cite{Celledoni14,Celledoni24}. 
For the bi-Hamiltonian formulation, the conservative Hamiltonian part is integrated by the implicit energy- preserving average vector field (AVF) method. 
We demonstrate in numerical experiments that the Hamiltonians are preserved for the conservative parts of the Lorenz, Chen, and Rabinovich systems. 
The periodic and chaotic solutions of these three dissipative bi-Hamiltonian/Nambu systems are computed accurately with the conservative-dissipative Strang splitting .

The outline of the paper is as follows.
In Section~\ref{sec:nambu}, we introduce the framework for the 3D dissipative bi-Hamiltonian/Nambu systems. 
We present Lorenz, Chen, and Rabinovich systems with non-dissipative and dissipative parts, Nambu Hamiltonians, dissipation functions, and the bi-Hamiltonian forms. 
In Section~\ref{sec:strang},  we  describe the structure-preserving Strang splitting method with Kahan's and AVF  methods. 
We demonstrate in Section~\ref{sec:numeric} the preservation of Hamiltonians, periodic and chaotic solutions of the Lorenz, Chen, and Rabinovich systems.  
Conclusions are given in the last section.

%%%%%%%%%%%%%%%%%%%%%%%%%%%%%%%%%%%%%%%%%%%%%%%%%%%%%%%%%%%%%%%%%%%%%%%%%%%%%%%%%%%%%%
%%%%%%%%%%%%%%%%%%%%%%%%%%%%%%%%%%%%%%%%%%%%%%%%%%%%%%%%%%%%%%%%%%%%%%%%%%%%%%%%%%%%%%
\section{3D dissipative bi-Hamiltonian/Nambu systems}
\label{sec:nambu}

In three dimensions, the Hamiltonian (reversible) part of a metriplectic system
can be written in the terms of bi-Hamiltonian/Nambu bracket \cite{Esen17}.
Nambu-Hamilton mechanics is a specific generalization of classical
Hamiltonian mechanics, where the invariance group of canonical
symplectic transformations of the
Hamiltonian evolution equations in $2n$ dimensional phase space are extended to the
more general
volume preserving transformation group \cite{Nambu73,Takhtajan94}.
Nambu-Hamiltonian form  of a particular dynamical system in $\mathbb{R}^{3}$ is defined by
two scalar functions, i.e., the generalized or Nambu
Hamiltonians $H_{1}, H_{2}\in C^{\infty}(\mathbb{R}^{3})$:
\begin{equation} \label{nambu1}
%\frac{d\bx_i}{dt}=\dot{\bx}_{i} \ = \  \{ x^{i} , H_{1} , H_{2} \} \ \ \ \ \ \ \ \  i=1,2,3
\dot{\bx}  =  \nabla H_{1} \times \nabla H_{2}=
\begin{pmatrix}
\partial_yH_1\partial_zH_2 - \partial_zH_1\partial_yH_2 \\
\partial_zH_1\partial_xH_2 - \partial_xH_1\partial_zH_2 \\
\partial_xH_1\partial_yH_2 - \partial_yH_1\partial_xH_2
\end{pmatrix},
\end{equation}
where $\dot{\bx}$ denotes the ordinary derivative with respect to the time variable $t$, $\partial_{\omega}$ is the partial derivative operator with respect to the component $\omega\in\{x,y,z\}$, and $\bx = (x,y,z)^T$. 

In Hamiltonian mechanics, only one conserved quantity appears in the equation, whereas two conserved quantities have equal status in Nambu mechanics.
The phase space trajectory is represented
as the intersection line of two surfaces based on the conserved quantities. Therefore, the solutions can be illustrated without explicitly solving the equations of motion.

The Nambu $3$-bracket is a generalization of the Poisson bracket in
Hamiltonian mechanics. It is  defined for an arbitrary function $F$ as
$$
\dot{F} = \{F, H_1,H_2  \} := \nabla  \cdot(\nabla H_1 \times \nabla H_2).
$$ 
It satisfies the so-called fundamental identity \cite{Takhtajan94},  which is a generalized and stronger version of the Jacobi identity. 
Because of the cross product operation, the Nambu bracket is antisymmetric and also trilinear, and that the vector field of the Nambu-Hamiltonian form \eqref{nambu1} is divergence-free:
$$
\nabla \cdot \dot{\bx} = \nabla \cdot (\nabla H_1 \times \nabla H_2) = 0,
$$
yielding that the flow vector field  $\mathbf{v} :=  \dot{\bx}$ is  volume preserving.

On the other hand, the  dissipative  Nambu-Hamilton system have the form \cite{Axenides10,Katagiri26,Roupas12}
\begin{equation} \label{nambu2}
\dot{\bx} \ \ = \ \ \nabla H_{1} \times \nabla H_{2}
+ \nabla D,
\end{equation}
where the dissipation function $D$ characterizes the dissipative term.
The phase-space volume-preserving flows are called "non-dissipative",
while the non-conserving ones are "dissipative", i.e.,  $\partial_{x_i} v_{i}(\bx) > 0$ and $\partial_{x_i} v_{i}(\bx) < 0$, respectively, for $i=1,2,3$ and $\{x_1,x_2,x_3\}=\{x,y,z\}$.  
The Nambu Hamiltonians $ H_{1}$ and $H_{2} $ are conserved in the non-dissipative part of the system \eqref{nambu2}, i.e., when it reduces to the system \eqref{nambu1}.

The vector field of the dissipative Nambu-Hamiltonian system \eqref{nambu2} splits into non-dissipative and dissipative parts as
\begin{equation} \label{flow}
\dot{\bx} = f(\bx) = f_{nd}(\bx) + f_{d}(\bx),
\end{equation}
where the functions $f_{nd}(\bx)$ and $f_{d}(\bx)$ refer to the non-dissipative and dissipative parts, respectively.
3D quadratic systems such as Lorenz, Rabinovich and Chen systems can be written in dissipative Nambu-Hamilton form \eqref{nambu2} with quadratic Nambu Hamiltonians.

The Nambu-Hamiltonian system \eqref{nambu1} can also be written in bi-Hamiltonian form for state dependent antisymmetric matrices $J_1(\bx)$ and $J_2(\bx)$ as follows  (see, for example, \cite{Esen17})
\begin{equation} \label{biham}
\dot{\bx} = J_1(\bx) \nabla H_1, \qquad \dot{\bx} = J_2(\bx) \nabla H_2
\end{equation}
for which 
\begin{equation} \label{biham1}
\nabla H_1 \times \nabla H_2=J_1\nabla H_1, \quad
J_1(\bx) = 
\begin{pmatrix}
0 & \partial_zH_2 & -\partial_yH_2 \\
-\partial_zH_2 & 0 & \partial_xH_2 \\
\partial_yH_2 & -\partial_xH_2 & 0
\end{pmatrix},
\end{equation}
and
\begin{equation} \label{biham2}
\nabla H_1 \times \nabla H_2=J_2\nabla H_2, \quad 
J_2(\bx) = 
\begin{pmatrix}
0 & -\partial_zH_1 & \partial_yH_1 \\
\partial_zH_1 & 0 & -\partial_xH_1 \\
-\partial_yH_1 & \partial_xH_1 & 0
\end{pmatrix}.
\end{equation}

Consequently, when the dissipation function $D$ is at most quadratic, the dissipative Nambu-Hamilton system \eqref{nambu2} can be recast as a linearly perturbed  dissipative system in bi-Hamiltonian  form as
\begin{equation}
\dot{\bx} = J_1(\bx) \nabla H_1 + \nabla D, \qquad \dot{\bx} = J_2(\bx) \nabla H_2 + \nabla D.
\end{equation}

In the following, we consider three representative dissipative bi-Hamiltonian/Nambu systems of the form \eqref{nambu2}.

%%%%%%%%%%%%%%%%%%%%%%%%%%%%%%%%%%%%%%%%%%%%%%%%%%%%%%%%%%%%%%%%%%%%%%%%%%%%%%%%%%%%%%
\subsection{Lorenz System}

The Lorenz system is widely known as a simplified model of convection and fluid dynamics \cite{Lorenz83}
\begin{align} \label{lorenz}
\dot{x} &= \sigma(y - x),   \nonumber\\
\dot{y} &= x(\rho-z) - y, \\
\dot{z} &= xy - \beta z. \nonumber
\end{align}
where $\sigma$ is the Prandtl number, $\rho$  is the relative Reynolds number and $\beta$ is the geometric aspect ratio.
The flow vector field of Lorenz system consists of non-dissipative and dissipative parts \cite{Axenides10}
\begin{align*}
 \nabla H_{1} \times \nabla H_{2}  = f_{nd}(\bx) &= ( \sigma y, x(\rho-z), xy)^T, \\
\nabla D = f_{d}(\bx)  &=  ( -\sigma x , -y , -\beta z )^T,
\end{align*}
with the Nambu  Hamiltonians and dissipation function 
\begin{align*}
H_{1}  &= \frac{1}{2} ( y^{2} + ( z- \rho)^{2} ), \quad H_{2}\ = \ \sigma z \ - \ \frac{x^{2}}{2}, \\
D  &=   - \frac{1}{2}  ( \sigma x^{2} \ + \ y^{2} \ + \ \beta z^{2}).
\end{align*}
%\begin{eqnarray*}
%\dot{H}_1 & = & -y^2 + \beta z(z -\rho), \quad  \dot{H}_2  =  \sigma x^2 -\sigma \beta z,\\ 
%\dot{D} & = & \sigma^2x^2 + y^2 + \beta^2 z^2 -(\sigma + \rho)xy  + (1-\beta)xyz.
%\end{eqnarray*}

The Lorenz system \eqref{lorenz} can be written in the linearly perturbed bi-Hamiltonian systems as: according to $\dot{\bx} = J_1(\bx) \nabla H_1 + \nabla D$ with $J_1$ in \eqref{biham1}

$$
\dot{\bx} = 
\begin{pmatrix}
0 & \sigma & 0 \\
-\sigma & 0 & -x \\
0 & x & 0
\end{pmatrix}
\begin{pmatrix}
0 \\
y \\
z-\rho
\end{pmatrix} - \begin{pmatrix}
\sigma x \\
y\\
\beta z
\end{pmatrix}, 
$$
or, according to $\dot{\bx} = J_2(\bx) \nabla H_2 + \nabla D$ with $J_2$ in \eqref{biham2}
$$
\dot{\bx} = 
\begin{pmatrix}
0 & \rho -z & y \\
z-\rho & 0 &0 \\
-y & 0 & 0
\end{pmatrix}
\begin{pmatrix}
-x \\
0\\
\sigma
\end{pmatrix} - \begin{pmatrix}
\sigma x\\
y \\
\beta z
\end{pmatrix}.
$$

%%%%%%%%%%%%%%%%%%%%%%%%%%%%%%%%%%%%%%%%%%%%%%%%%%%%%%%%%%%%%%%%%%%%%%%%%%%%%%%%%%%%%%
\subsection{Chen system}
The Chen system \cite{Chen99} is a representative dynamical system with chaotic behavior similar to the Lorenz system. The Chen system exhibits a parameter-dependent contraction rate, with continuous deformations of the phase-space structure.
Unlike the Lorenz system where the phase-space contraction rate is constant, bifurcations in Chen system  produce abrupt changes in the contraction rate.
For the real parameters $a$, $b$ and $c$, the Chen system consists of the following set of three ordinary differential equations:
\begin{align} \label{chen}
\dot{x} &= a(y - x), \nonumber\\
\dot{y} &= (c-a)x+cy-xz, \\
\dot{z} &= xy - b z. \nonumber
\end{align}
%where the parameters are set as $a = 35$, $b = 3$ w. %While the Lorenz system exhibits a constant rate of phase-space volume contraction (uniformly dissipative), the Chen system's contraction rate varies with the parameter $c$, resulting in a continuous change.
The non-dissipative and dissipative parts of the Chen system \eqref{chen} are 
\begin{align*}
f_{nd} & =    ( (2a-c)y, -xz , xy )^T, \\ 
f_{d}  & =  ( -ax + (c-a)y, cy  + (c-a)x , -bz)^T,
\end{align*}
with the Nambu Hamiltonians and dissipation function are given by \cite{Katagiri26}
\begin{align*}
H_1 &=  \frac{1}{2}(2a-c)(y^2+z^2), \quad
H_2 = -z + \frac{1}{2(2a-c)}x^2,\\
D  &=   -\frac{1}{2}ax^2 + \frac{1}{2}cy^2 - \frac{1}{2}bz^2 + (c-a) xy.
\end{align*}

The Chen system \eqref{chen} can be written in the linearly perturbed bi-Hamiltonian systems as: according to $\dot{\bx} = J_1(\bx) \nabla H_1 + \nabla D$ with $J_1$ in \eqref{biham1}

$$
\dot{\bx} = 
\begin{pmatrix}
0 & -1 & 0 \\
1 & 0 & \frac{1}{2a-c} \\
0 & - \frac{1}{2a-c} & 0
\end{pmatrix}
\begin{pmatrix}
0 \\
(2a -c)y \\
(2a -c)z 
\end{pmatrix} + \begin{pmatrix}
-\alpha x + (c-a) y \\
(c-a) x + c y \\
-bz
\end{pmatrix}
$$
or, according to $\dot{\bx} = J_2(\bx) \nabla H_2 + \nabla D$ with $J_2$ in \eqref{biham2}

$$
\dot{\bx} = 
\begin{pmatrix}
0 & (2a -c)z  & (c-2a)y \\
(c-2a)z & 0 &0 \\
(2a -c)y & 0 & 0
\end{pmatrix}
\begin{pmatrix}
\frac{1}{2a-c} x \\
0\\
-z
\end{pmatrix} + \begin{pmatrix}
-\alpha x + (c-a) y \\
(c-a) x + c y \\
-bz
\end{pmatrix}.
$$

%%%%%%%%%%%%%%%%%%%%%%%%%%%%%%%%%%%%%%%%%%%%%%%%%%%%%%%%%%%%%%%%%%%%%%%%%%%%%%
\subsection{Rabinovich system}

The Rabinovich system  \cite{Rabinovich81,Libres08,Giacomini12,Ghosh25}  is described by the following set of ordinary differential equations  
\begin{align}\label{rab}
\dot{x}  & = qy -k_1x  + y z, \nonumber \\
\dot{y} & =  q x - k_2 y -x z,  \\
\dot{z} & =  -k_3z  + x y,\nonumber 
\end{align}
where  $k_1,k_2,k_3$ are the damping rates. It is a dynamical system
of three resonantly coupled waves, parametrically excited.
values for which a strange attractor similar to the Lorenz equation is produced.

The Nambu-Hamilton formulation of the Rabinovich system \eqref{rab} was derived in \cite{Esen17} with the non-dissipative and dissipative parts, respectively,
$$
f_{nd}  =   (y z,-x z,x y )^T, \quad
f_{d}   =  (qy-k_1 x, qx -k_2y,-k_3z  )^T,
$$
and with the  Nambu Hamiltonians and the dissipation function 
\begin{align*}
H_{1}  & =  \frac{1}{2} \left ( x^2 +  y^2 \right), \quad  H_{2}  =  \frac{1}{2} \left ( y^2 + z^2\right),\\ 
D & = - \frac{1}{2} \left (k_1 x^2  -qx^2 +k_2y^2 -qy^2-k_3z^2\right). 
\end{align*}

The Rabinovich system \eqref{rab} can be written in the linearly perturbed bi-Hamiltonian systems as: according to $\dot{\bx} = J_1(\bx) \nabla H_1 + \nabla D$ with $J_1$ in \eqref{biham1}

$$
\dot{\bx} = 
\begin{pmatrix}
0 & 0 & -y \\
0 & 0 & x \\
y & -x & 0
\end{pmatrix}
\begin{pmatrix}
0 \\
y \\
z
\end{pmatrix} + 
\begin{pmatrix}
qy -k_1x\\
k_2 y -qy \\
-k_3 z
\end{pmatrix},
$$
or, according to $\dot{\bx} = J_2(\bx) \nabla H_2 + \nabla D$ with $J_2$ in \eqref{biham2}

$$
\dot{\bx} = 
\begin{pmatrix}
0 & z & -y \\
-z & 0 &0 \\
y & 0 & 0
\end{pmatrix}
\begin{pmatrix}
x \\
y\\
0
\end{pmatrix} + \begin{pmatrix}
qy -k_1x\\
k_2 y -qy \\
-k_3 z
\end{pmatrix}.
$$

%%%%%%%%%%%%%%%%%%%%%%%%%%%%%%%%%%%%%%%%%%%%%%%%%%%%%%%%%%%%%%%%%%%%%%%%%%%%%%%%%%%%%
%%%%%%%%%%%%%%%%%%%%%%%%%%%%%%%%%%%%%%%%%%%%%%%%%%%%%%%%%%%%%%%%%%%%%%%%%%%%%%%%%%%%
\section{Structure-preserving splitting integration}
\label{sec:strang}

Strang splitting is a second-order time integration method that employs symmetric operator splitting to solve nonlinear ordinary and partial differential equations.
The method decomposes operators so that sub-flows are integrated individually.

%%%%%%%%%%%%%%%%%%%%%%%%%%%%%%%%%%%%%%%%%%%%%%%%%%%%%%%%%%%%%%%%%%%%%%%%%%%%%%%%%%%%%%
\subsection{Splitting integrator of dissipative Nambu systems with Kahan's method}

Consider the one-step numerical scheme $\bx_{k+1} = \Phi_{h} \bx_k$ for the solution of the dissipative Nambu system \eqref{flow}, where for the step size $h$, the map $\Phi_{h}$ is the composition \cite{McLachlan02s,Blanes24}
\begin{equation} \label{strang}
\Phi_{h} = \varphi\left(\frac{h}{2} f_d \right) \circ \varphi \left( h f_{nd} \right) \circ \varphi \left(\frac{h}{2} f_{d}\right),
\end{equation}
with some numerical flow map $\varphi$.
The divergence-free  non-dissipative part is integrated with the second order, time-reversal Kahan's  method  \cite{Celledoni13,Kahan93,Kahan97}, and the dissipative part by forward and backward Euler methods,  so that the composition is symmetric, i.e.,
$\Phi_{-h}\circ\Phi_{ h}=I$.

Kahan's method  is designed as an integrable discretization of linear-quadratic systems:
$$
\dot{\bx} = f(\bx) =  Q(\bx) + L{\bx},
$$
leading to the scheme
\begin{equation}\label{kahan1}
\frac{\bx^{k+1} - \bx^k}{h} = \widetilde{Q}(\bx^k,\bx^{k+1}) + \frac{1}{2}L(\bx^k + \bx^{k+1}),
\end{equation}
where, the symmetric bilinear form $\widetilde{Q}(\cdot ,\cdot )$ is obtained by the polarization of the quadratic vector field $Q(\cdot)$ 
\begin{equation*}
\widetilde{Q}(\bx^k,\bx^{k+1}) := \frac{1}{2}\left( Q(\bx^k+\bx^{k+1}) - Q(\bx^k) -  Q(\bx^{k+1}) \right).
\end{equation*}
Kahan's method coincides with the Runge-Kutta method restricted to quadratic vector fields \cite{Celledoni13}
\begin{equation} \label{kahan}
\frac{\bx^{k+1} - \bx^k}{h} = -\frac{1}{2}f(\bx^{k+1} )  +  2 f\left( \frac{\bx^{k+1} + \bx^k }{2}  \right)   -\frac{1}{2}f(\bx^{k} ).
\end{equation}

Moreover, it is linearly-implicit; the solution  $\bx^{k+1}$  of  \eqref{kahan}
is computed by solving a single linear system of equations
$$
\left ( I -\frac{h}{2} f'(\bx^k)\right ) \widetilde{\bx} = h \bf(\bx^k),\qquad \bx^{k+1} = \bx^k + \widetilde{\bx},
$$
where $f'$ denotes the Jacobian matrix of $f$.

Kahan's method is symmetric, and hence it is second-order  \cite{Celledoni13}.
 For Hamiltonian systems with quadratic vector fields, it preserves the modified volume and modified Hamiltonians \cite{Celledoni14,Iserles15}.
The homogeneous and inhomogeneous Nambu Hamiltonians of the form 
$$
H_1(\bx) = \bx^T A  \bx + a^T \bx, \quad  H_2(\bx) = \bx^T B \bx + b^T\bx,
$$
are preserved, where $A$, and $B$ are arbitrary $3\times 3$ matrices and $a$ and $b$ are vectors in $R^3$ \cite{Celledoni19,Celledoni24}. The Nambu Hamiltonians  of the Lorenz, Chen, and  Rabinovich systems are in this form.

%%%%%%%%%%%%%%%%%%%%%%%%%%%%%%%%%%%%%%%%%%%%%%%%%%%%%%%%%%%%%%%%%%%%%%%%%%%%%%%%%%%%%%
\subsection{Splitting integrator of dissipative bi-Hamiltonian  systems with AVF  method}

For dissipative bi-Hamiltonian systems \eqref{biham}, the conservative Hamiltonian part is integrated by the energy preserving average vector field (AVF) method \cite{Quispel08}, and the linearly perturbed dissipative parts, by forward and backward Euler methods. For a Poisson system like \eqref{biham}
$$
\dot{\bx} = J(\bx) \nabla H,
$$
the AVF method  is given as \cite{Cohen11,Karasozen13}
\begin{equation} \label{avf}
\frac{\bx^{k+1} -\bx^k}{h} = J\left (  \frac{\bx^{k+1} +\bx^k}{2} \right ) \int_0^1 \nabla H (\bx^k + \tau (\bx^{k+1} -\bx^k))d\tau.
\end{equation}

The AVF method preserves the quadratic Hamiltonians and Casimirs of  Poisson systems \cite{Cohen11} such as  the conservative parts of  Lorenz, Chen, and Rabinovich systems.  In contrast to  Kahan's method, the AVF method is implicit and  not volume-preserving. 

\begin{remark} 
Exponential integrators based on discrete gradients  including the AVF method  \cite{Moore21} and Kahan's method \cite{Karasozen26lie} are applied to linearly damped Poisson systems
under the assumption that the linear dissipative part has a constant diagonal coefficient matrix, which is only the case for the Lorenz equation. 
\end{remark}

%%%%%%%%%%%%%%%%%%%%%%%%%%%%%%%%%%%%%%%%%%%%%%%%%%%%%%%%%%%%%%%%%%%%%%%%%%%%%%%
%%%%%%%%%%%%%%%%%%%%%%%%%%%%%%%%%%%%%%%%%%%%%%%%%%%%%%%%%%%%%%%%%%%%%%%%%%%%%%%
\section{Numerical results}
\label{sec:numeric}

In this section, we present numerical results for the dissipative bi-Hamiltonian/Nambu systems introduced in Section~\ref{sec:nambu}, i.e., Lorenz, Chen and Rabinovich systems. The systems are solved by the Strang splitting \eqref{strang} using the step size $h=0.001$. For the conservative parts of the systems, Kahan's method is used in Nambu form, while we use AVF method when the system is in bi-Hamiltonian form. For the system in bi-Hamiltonian form, we give the results for only the Lorenz system, as the results for the other problems are similar. In either problem, we present the phase plots for the systems and the ones for the non-dissipative parts. We also demonstrate the Hamiltonian errors for the non-dissipative parts by plotting the relative errors
$$
\frac{H_i^k - H_i^0}{H_i^0}, \quad i=,1,2,
$$ 
where $H_i^k$ denotes the value of Hamiltonian $H_i$ at time $t_k=hk$.

%%%%%%%%%%%%%%%%%%%%%%%%%%%%%%%%%%%%%%%%%%%%%%%%%%%%%%%%%%%%%%%%%%%%%%%%%%%%%%%
\subsection{Lorenz system}

We consider the Lorenz system \eqref{lorenz} with the standard parameter values $\sigma = 10$, $\beta = 8/3$ and $\rho = 14 $. 
We prescribe the initial conditions as $x(0)=y(0)=z(0)=1$, and the terminal time is $t=20$.

\begin{figure}[htb!]
\centering
\includegraphics[width=0.28\columnwidth]{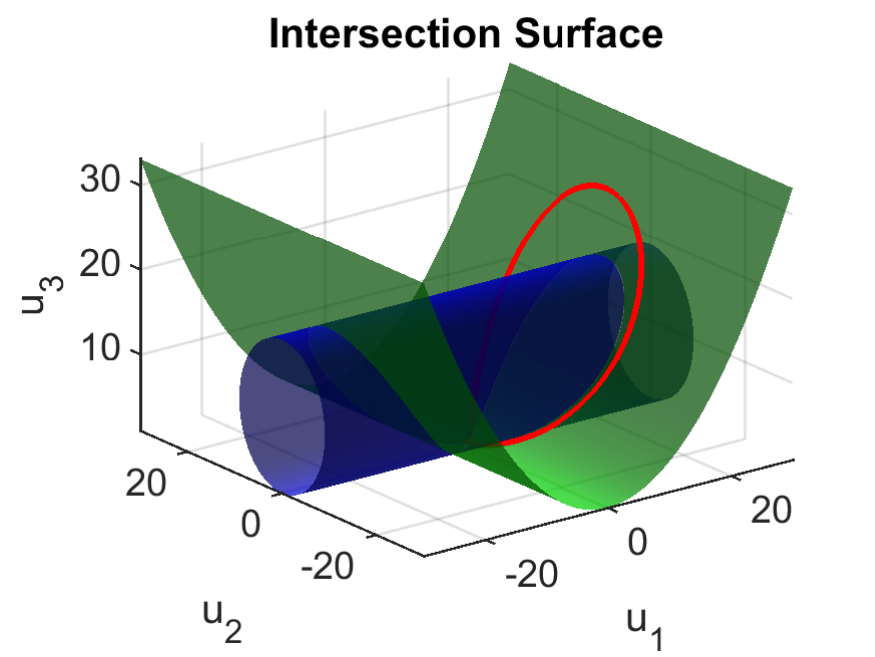}
\includegraphics[width=0.28\columnwidth]{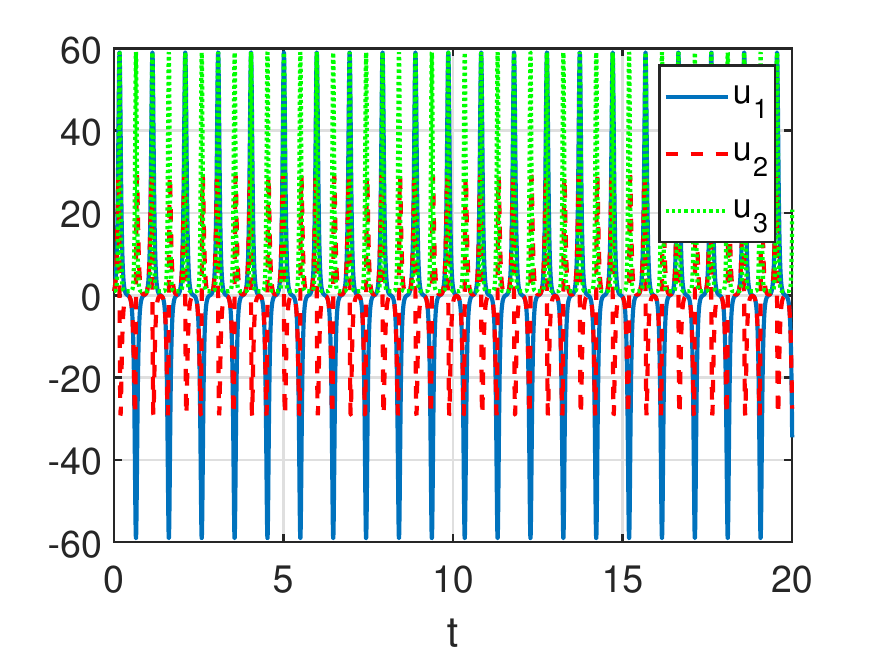}
\includegraphics[width=0.28\columnwidth]{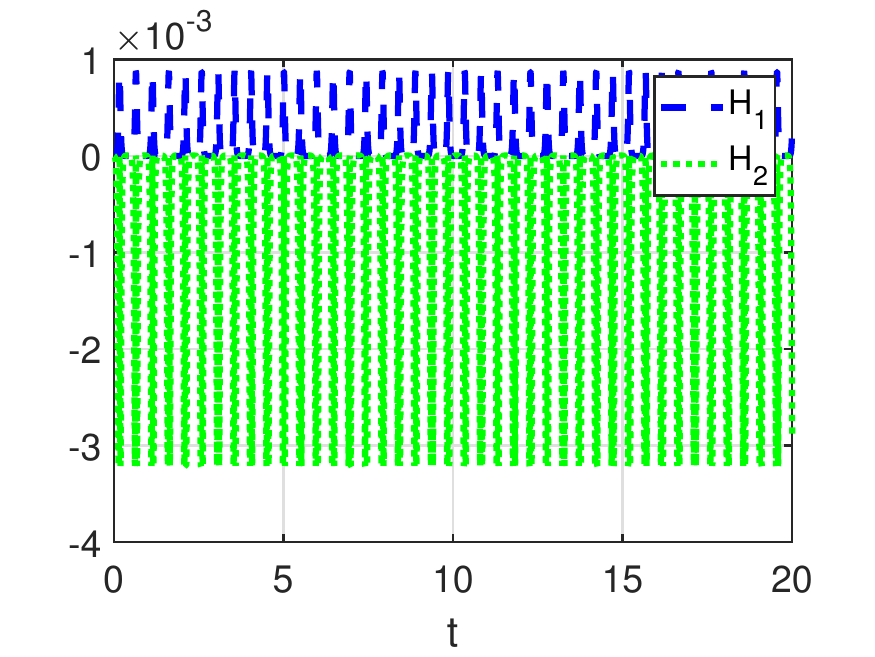}
\caption{Lorenz equation: non-dissipative part solved by Kahan's method for $\rho =14$.}
\label{lork}
\end{figure}

For the Lorenz system, the intersection surface calculated directly from the Nambu Hamiltonians is presented in \figurename~\ref{lork}, left. In that figure, we also give the solution trajectories and the relative Hamiltonian errors for the non-dissipative part of the Lorenz system in Nambu form, solved by Kahan's method. We observe that the Hamiltonian errors oscillate and are bounded over time, demonstrating the preservation of the Hamiltonians.

\begin{figure}[htb!]
\centering
\includegraphics[width=0.28\columnwidth]{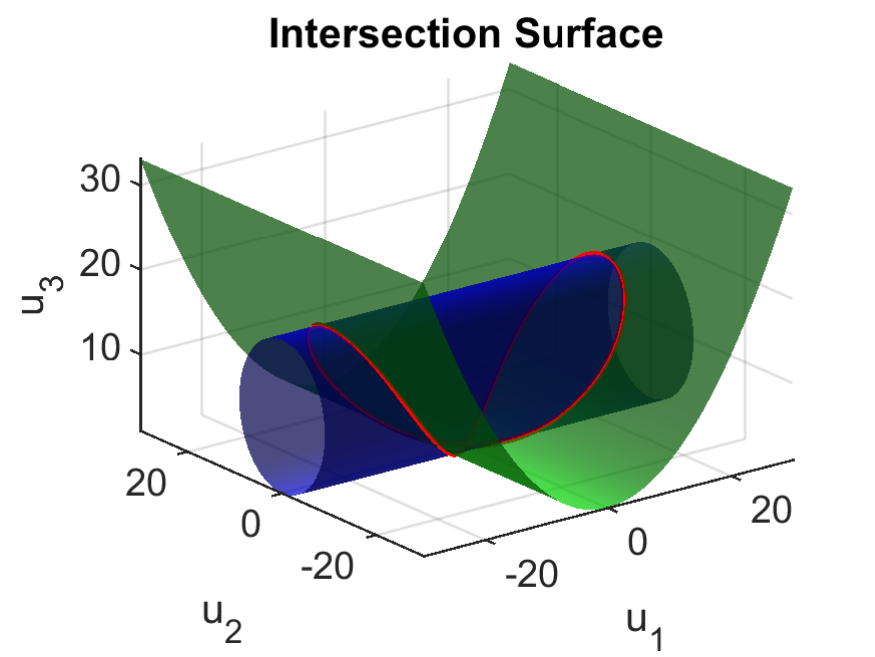}
\includegraphics[width=0.28\columnwidth]{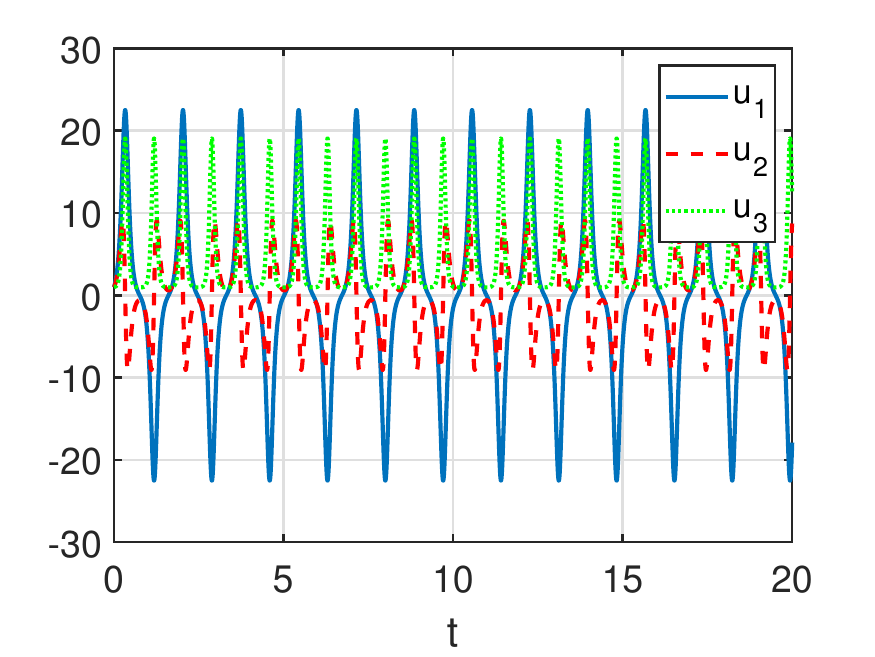}
\includegraphics[width=0.28\columnwidth]{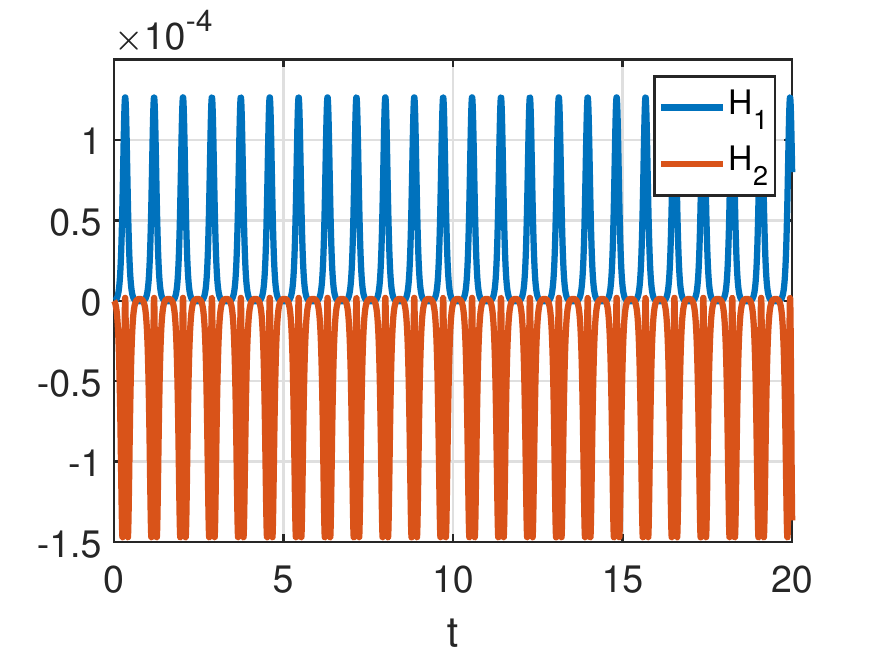}
\caption{Lorenz equation: non-dissipative part solved by AVF method for $\rho =14$.}
\label{lorm}
\end{figure}

In \figurename~\ref{lorm}, we present the same plots for the Lorenz system in bi-Hamiltonian form, and obtained by the AVF method \eqref{avf} as the integrator, which reduces to the mid-point rule because of the quadratic Hamiltonians. Similar to the results in \figurename~\ref{lork}, the Hamiltonians are conserved, as well.

\begin{figure}[htb!]
\centering
\includegraphics[width=0.28\columnwidth]{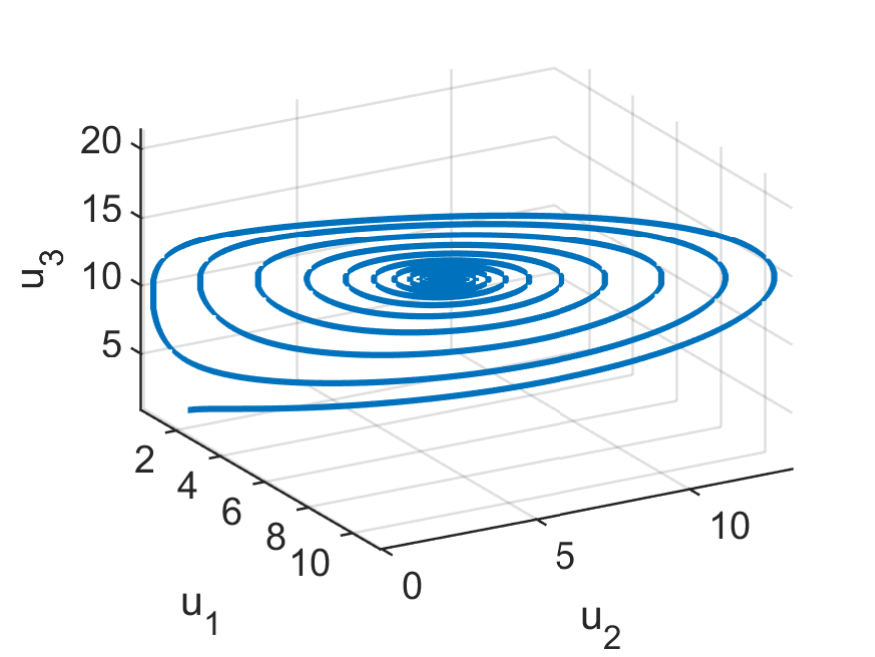}
\includegraphics[width=0.28\columnwidth]{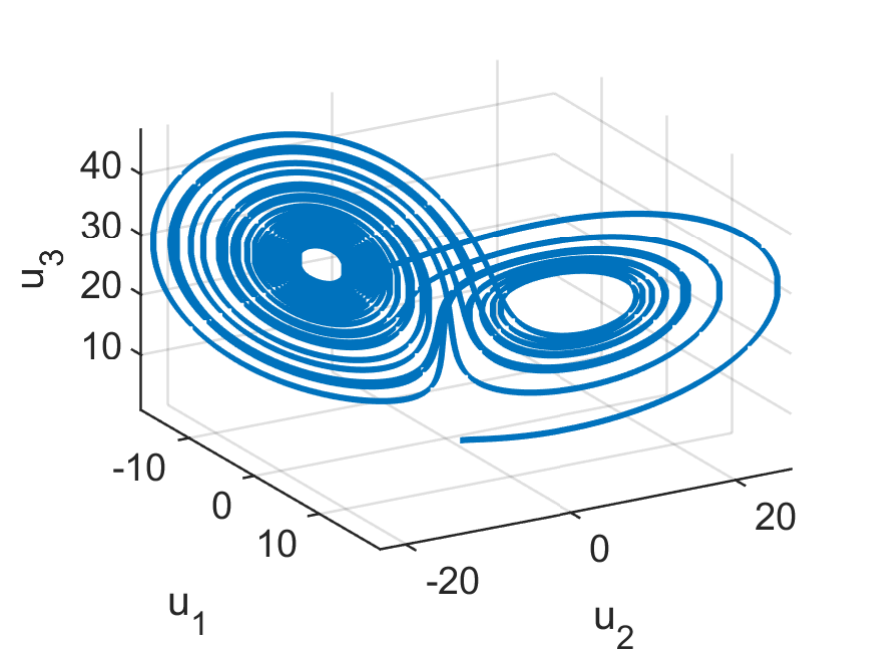}
\includegraphics[width=0.28\columnwidth]{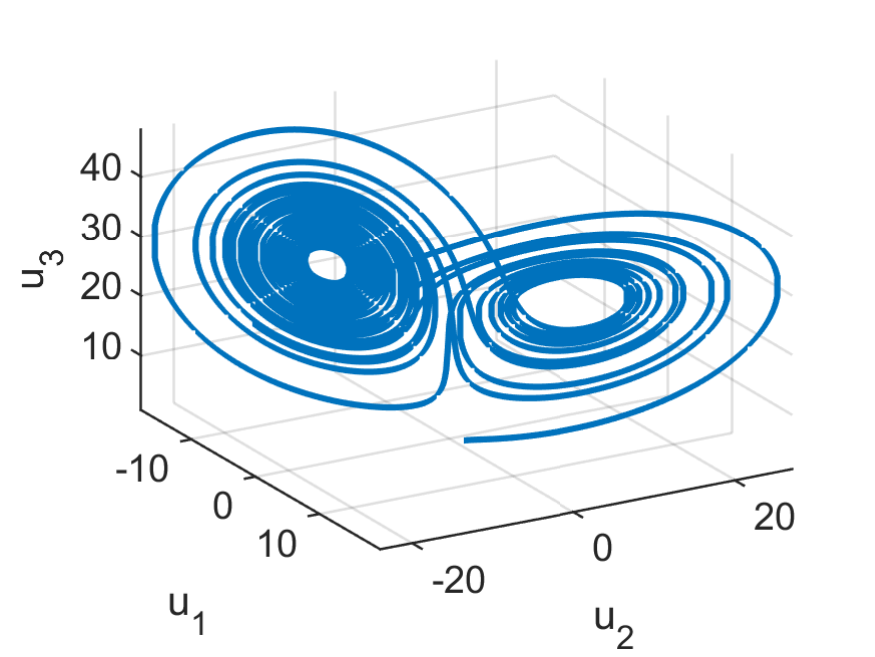}
\caption{Lorenz system: phase plots for  $\rho =14,24.74,28$ (from left to right).}
\label{lor2}
\end{figure}

In case of the parameter $\rho$, the Lorenz system has a nonzero stable equilibrium when $\rho < 24.74$, with trajectories eventually converging to it after damped oscillations.
For $\rho \approx 24.74$ a bifurcation occurs, leading to the emergence of a stable periodic solution.
When $\rho > 24.74 $, the periodic solution becomes unstable, and the system transitions to chaos. We give the phase plots for varying values of $\rho$ in \figurename~\ref{lor2}.

%\begin{figure}[htb!]
%\centering
%\includegraphics[scale=0.30]{lordiff}
%\caption{Lorenz equation: }
%\label{lordif}
%\end{figure}

%%%%%%%%%%%%%%%%%%%%%%%%%%%%%%%%%%%%%%%%%%%%%%%%%%%%%%%%%%%%%%%%%%%%%%%%%%%%%%%%%%%%%%
\subsection{Chen system}

We consider the Chen system \eqref{chen} with the fixed parameter values $a = 35$ and $b = 3$ \cite{Katagiri26}. 
The initial conditions are $x(0)=y(0)=z(0)=1$, and the terminal time is $t=20$.

For the parameter value $c=20$, the intersection surface of the Chen system is presented in \figurename~\ref{lork}, left. In that figure, we also give the solution trajectories and the relative Hamiltonian errors for the non-dissipative part of the Chen system in Nambu form, solved by Kahan's method. We observe that the Hamiltonians are preserved, similar to the Lorenz system.

\begin{figure}[ht!]
\centering
\includegraphics[width=0.28\columnwidth]{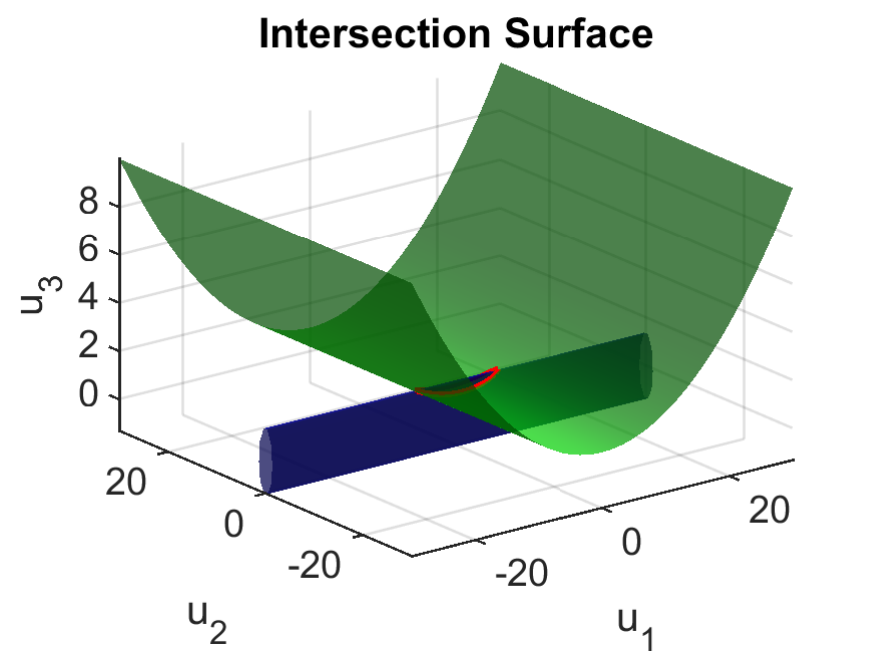}
\includegraphics[width=0.28\columnwidth]{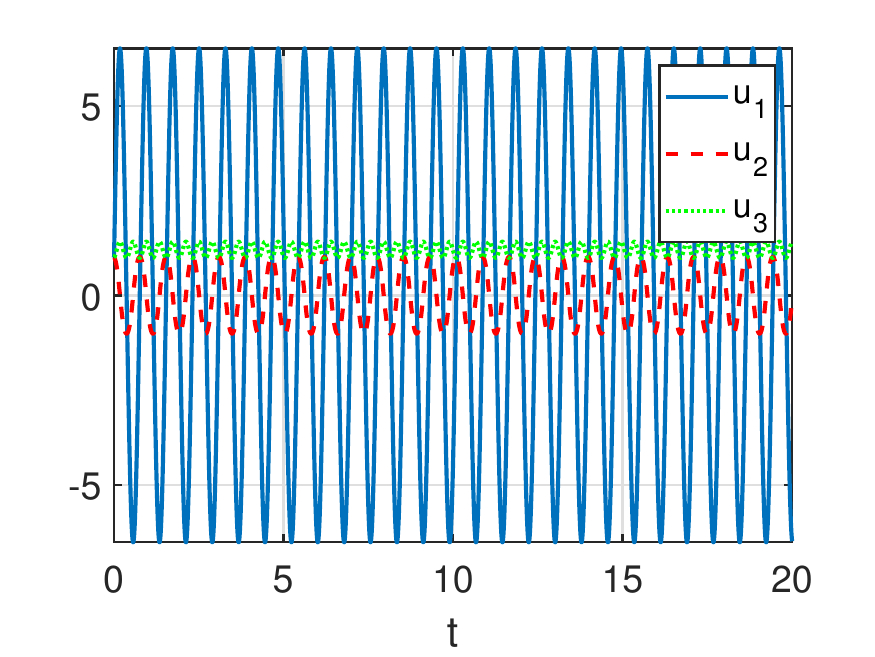}
\includegraphics[width=0.28\columnwidth]{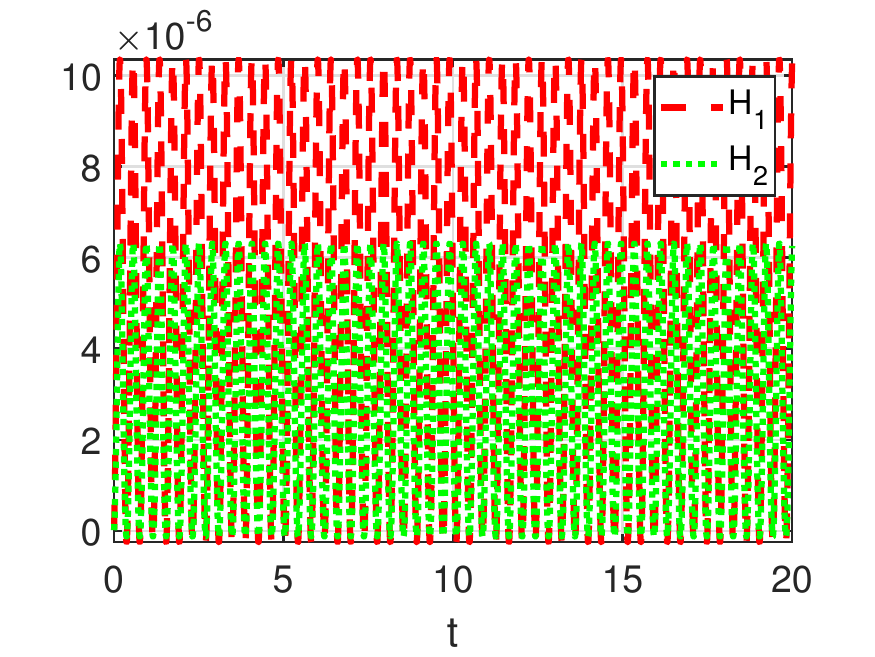}
\caption{Chen system: non-dissipative part.}
\label{chen1}
\end{figure}

\begin{figure}[ht!]
\centering
\includegraphics[width=0.28\columnwidth]{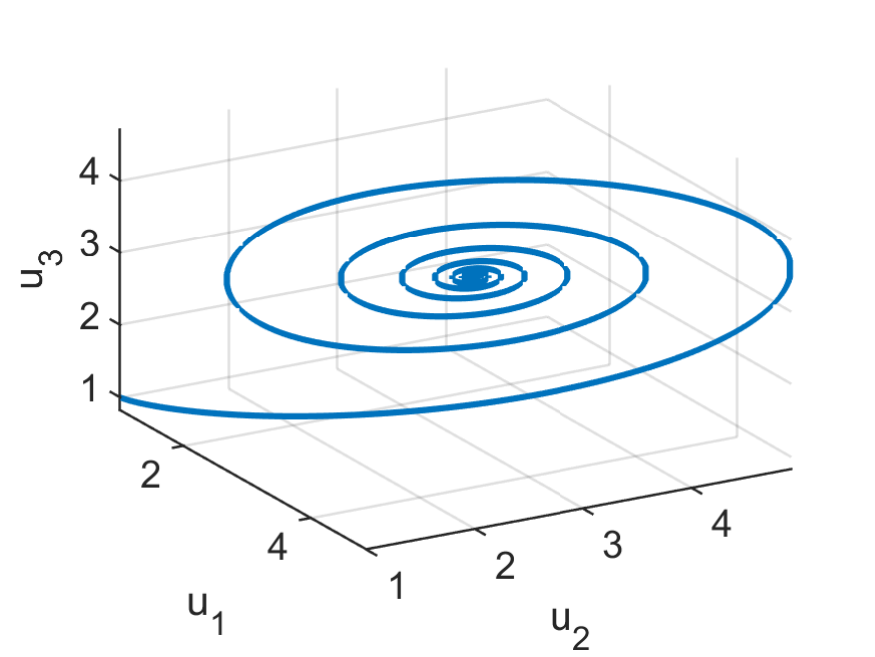}
\includegraphics[width=0.28\columnwidth]{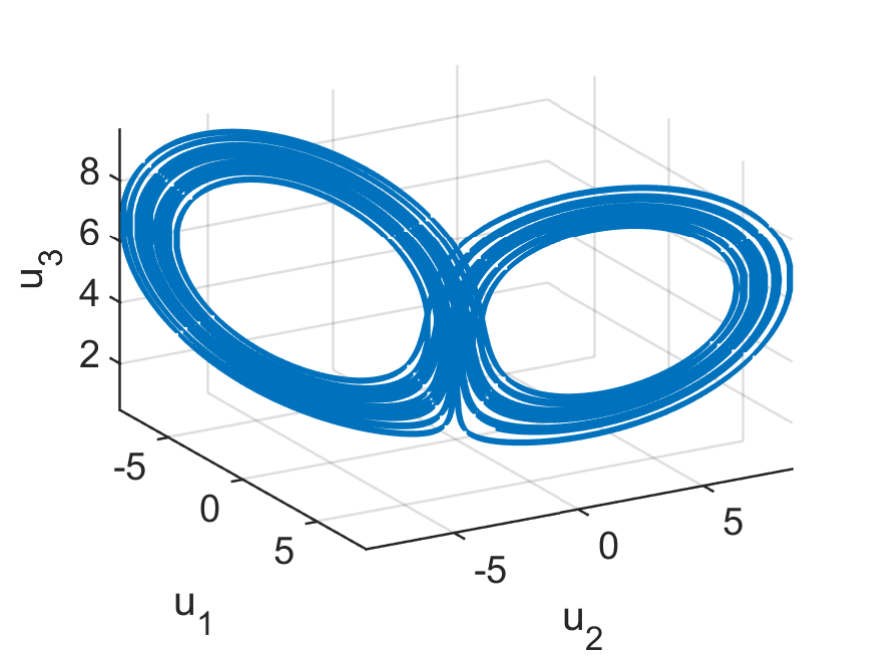}
\includegraphics[width=0.28\columnwidth]{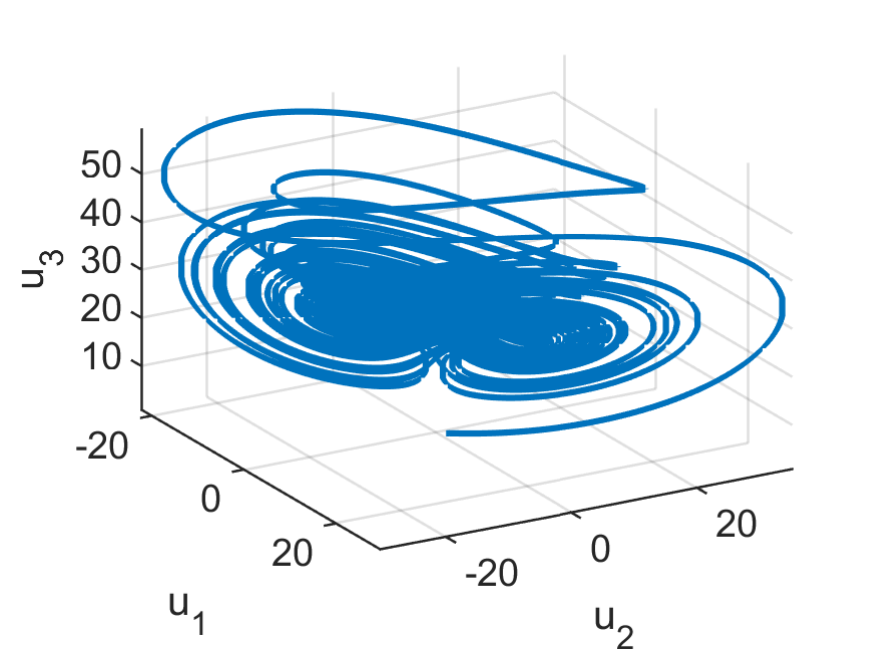}
\caption{Chen system: from left to right  $c=19,20.08,28$.}
\label{chen2}
\end{figure}

In  \figurename~\ref{chen2}, the trajectories converge to the stable equilibrium after damped oscillations when $17.5 < c < 20.08$.  
For $c \approx 20.08$, a bifurcation occurs, leading to the emergence of a stable periodic solution.
When $20.08 < c < 38$, the periodic solution becomes unstable, and chaos occurs as in \cite{Katagiri26}.

%\begin{figure}[htb!]
%\centering
%\includegraphics[scale=0.30]{chendiff}
%\caption{Lorenz equation: }
%\label{chendif}
%\end{figure}

%%%%%%%%%%%%%%%%%%%%%%%%%%%%%%%%%%%%%%%%%%%%%%%%%%%%%%%%%%%%%%%%%%%%%%%%%%%%%%%%%%%%%%
\subsection{Rabinovich  system}

We consider the Rabinovich system \eqref{rab} with the fixed parameter values $q =6.76$, $k_1=4$, and $k_2=k_3=1$. 
The initial conditions are $x(0)=1.5$, $y(0)=-1.25$ and $z(0)=3.5$, and the terminal time is $t=100$.

\begin{figure}[ht!]
\centering
\includegraphics[width=0.28\columnwidth]{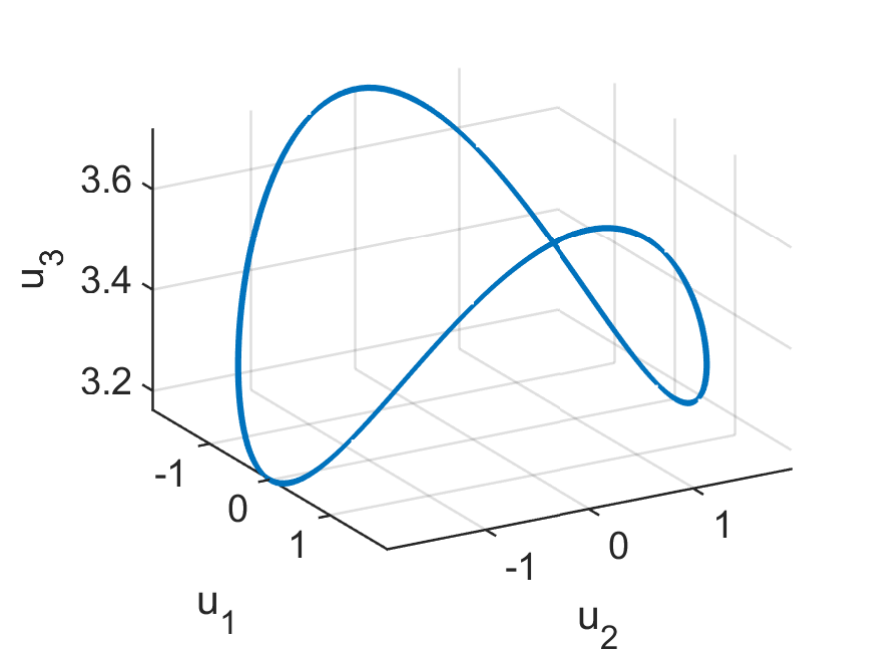}
\includegraphics[width=0.28\columnwidth]{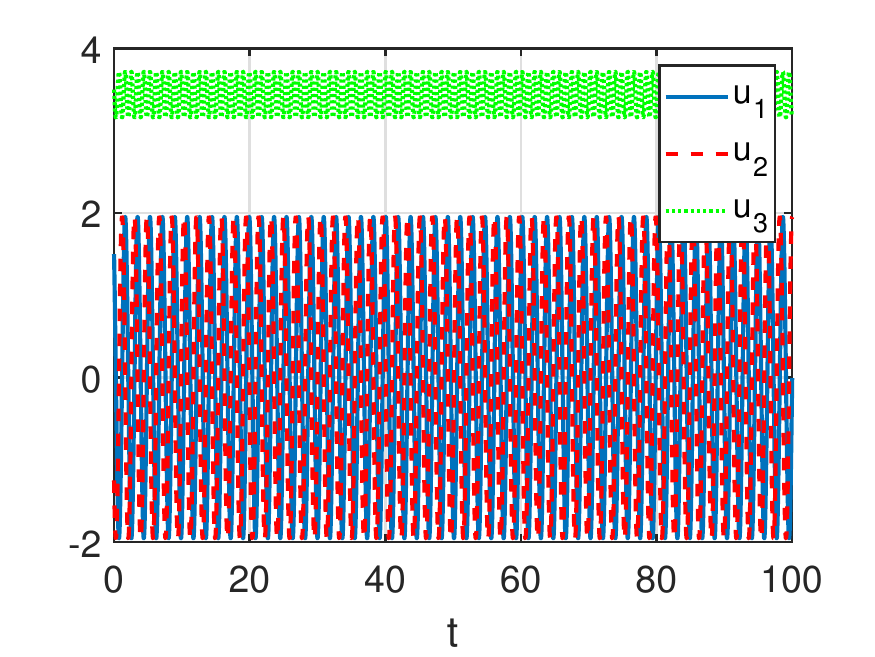}
\includegraphics[width=0.28\columnwidth]{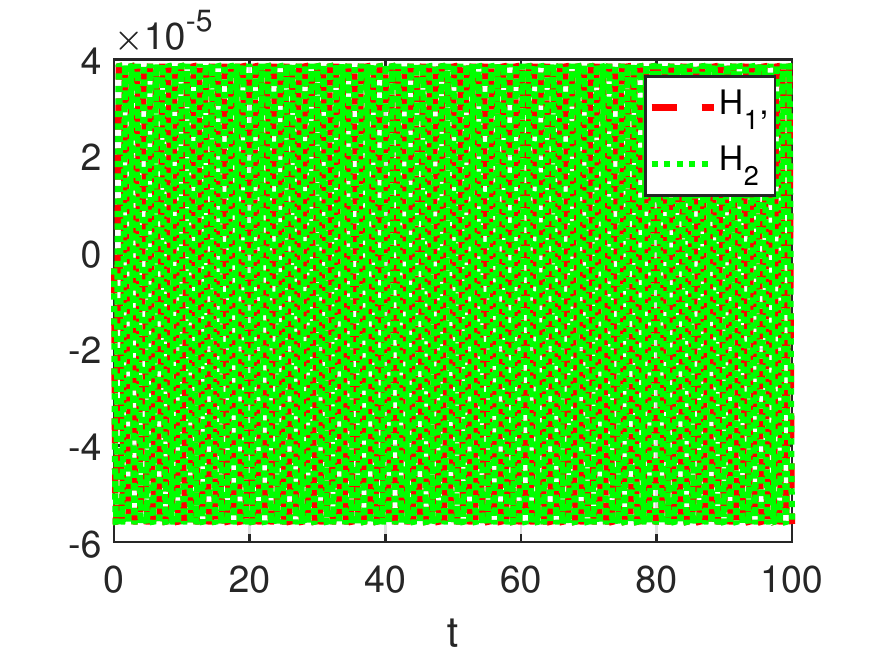}
\caption{Rabinovich system: non-dissipative part }
\label{rab1}
\end{figure}

\begin{figure}[ht!]
\centering
\includegraphics[width=0.35\columnwidth]{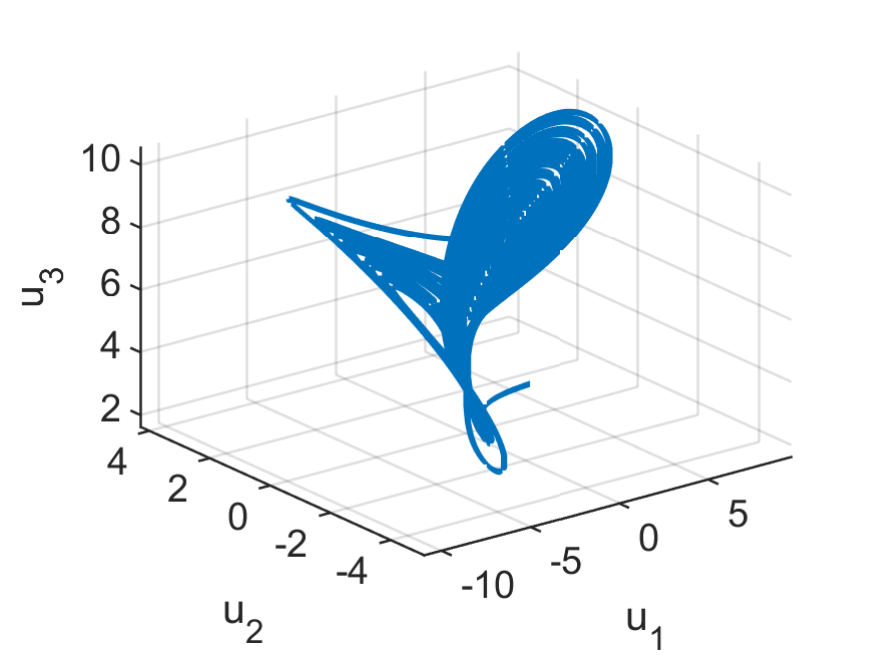}
\includegraphics[width=0.35\columnwidth]{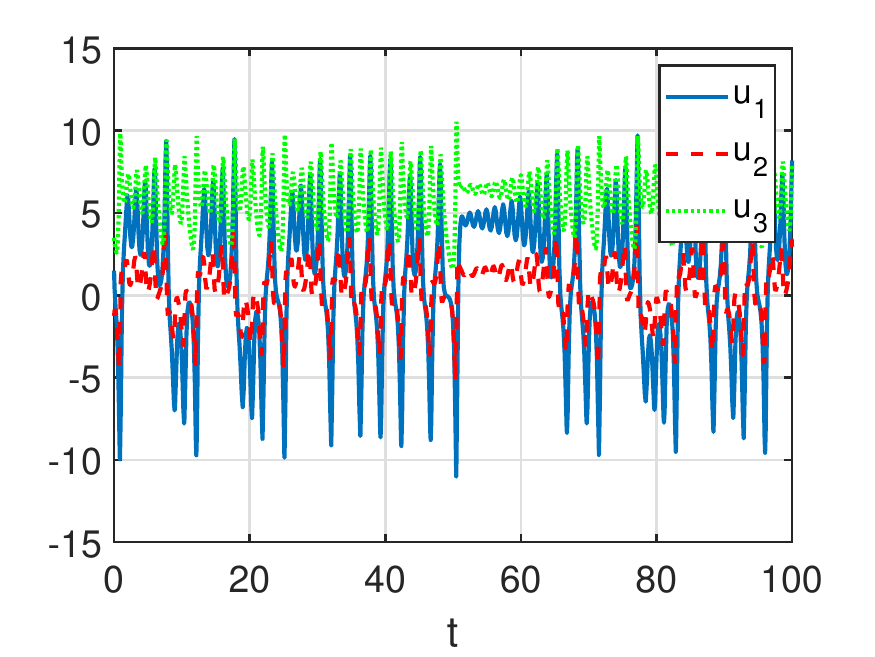}
\caption{Rabinovich system}
\label{rab2}
\end{figure}

The chaotic solution of the non-dissipative part of the Rabinovich system is shown in \figurename~\ref{rab1}, left. In \figurename~\ref{rab1}, right, the preservation of the Nambu Hamiltonians is illustrated with the oscillating solutions in \figurename~\ref{rab1}, middle.
We present the phase plot and solutions of the dissipative Rabinovich system in \figurename~\ref{rab2}.

%\cite{Libres08}
%$$
%x(0) = 5,  y(0) = 5, z(0) =  5, \quad  q =  0.006, k_1=4.5, k_2=-0.3, k_3=1.67.
%$$

%\begin{figure}[htb!]
%\centering
%\includegraphics[scale=0.30]{rabdiff}
%\caption{Lorenz equation: }
%\label{rabdif}
%\end{figure}

%%%%%%%%%%%%%%%%%%%%%%%%%%%%%%%%%%%%%%%%%%%%%%%%%%%%%%%%%%%%%%%%%%%%%%%%%%%%%%%%%%%%%%
%%%%%%%%%%%%%%%%%%%%%%%%%%%%%%%%%%%%%%%%%%%%%%%%%%%%%%%%%%%%%%%%%%%%%%%%%%%%%%%%%%%%%% 
\section{Conclusion} 	
	
A Structure-preserving conservative-dissipative Strang splitting is constructed for the dissipative bi-Hamiltonian/Nambu systems.  
Numerical results demonstrate that the Hamiltonians  of the conservative parts of the Lorenz, Chen and Rabinovich systems are preserved by Kahan's method and by the AVF method. 
Separating the conservative part from the dissipative part by splitting enables fast and accurate computation of the trajectories. 
This approach can be applied to 3D systems with generalized Hamiltonians \cite{Sarasola04}.

%\bibliographystyle{plain}
%\bibliography{references}

\end{document}